\documentstyle[aps,prb,psfig,epsfig,floats]{revtex}
\begin{document}
\draft
\twocolumn[\hsize\textwidth\columnwidth\hsize
\csname @twocolumnfalse\endcsname
\title{Renormalization  of the elementary
excitations in hole- and electron-doped cuprates due to 
spin fluctuations}
\author{D. Manske$^{1}$, I. Eremin$^{1,2}$, and K.H. Bennemann$^1$}
\address{$^1$Institut f\"ur Theoretische Physik, Freie Universit\"at 
Berlin, D-14195 Berlin, Germany}
\address{$^2$Physics Department, Kazan State University,
420008 Kazan, Russia}
\date{\today}
\maketitle
\begin{abstract}
  Extending our previous studies we present results for the doping-,
  momentum-, frequency-, and temperature- dependence of the
  kink-like change of the quasiparticle velocity resulting from the
  coupling to spin fluctuations.  In the nodal direction a kink is
  found in both the normal and superconducting state while in the
  antinodal direction a kink occurs only below $T_c$ due to the
  opening of the superconducting gap.  A pronounced kink is obtained
  only for hole-doped, but not for electron-doped cuprates and is
  characteristically different from what is expected due to
  electron-phonon interaction.  We further demonstrate that the kink
  structure is intimately connected to the resonance peak seen in
  inelastic neutron scattering.  Our results suggest similar effects
  in other unconventional superconductors like
  $\mbox{Sr}_2\mbox{RuO}_4$.

\end{abstract}
]
\narrowtext
\section{Introduction}
Elementary excitations in the cuprates are of central interest in
order to learn more about the correlations and the pairing mechanism
for superconductivity. For example, it is well-known that the
understanding of the elementary excitations in conventional
superconductors like lead as measured by tunneling spectroscopy played
the crucial role in accepting the picture of electron-phonon-mediated
Cooper-pairing \cite{bcs,wilki,rowell}. In the high-$T_c$ cuprates one
expects that due to the presence of antiferromagnetic spin
fluctuations a strong renormalization of the spectral density and the
corresponding energy dispersion may occur. It was shown by several
groups \cite{dahmchik,normanchik} that the so-called 'dip-hump'
structure seen in tunneling experiments \cite{zaza} can be explained
in terms of the feedback effect of superconductivity arising from the
structure in the gap function $\Delta({\bf k},\omega)$.  Moreover, it
has been argued that this structure reflects the effective pairing
interaction and points towards a spin-fluctuation-mediated
Cooper-pairing.  Recent developments in angle-resolved photoemission
spectroscopy (ARPES) allow to study the elementary excitations
directly and in much more detail. In particular, the anisotropy of the
elementary excitations close to the Fermi energy in the different
parts of the Brillouin Zone (BZ) are studied.  This is important,
since the coupling of the quasiparticles to spin fluctuations varies
at different parts of the BZ.  Furthermore, the understanding of the
structures seen by ARPES and their doping, momentum, and temperature
dependence will help to understand the role played by spin
fluctuations in contrast to phonons regarding the formation of
superconductivity in the high-$T_c$ cuprates.

\begin{figure}[h]
\vspace{0.5cm}
\centerline{\epsfig{file=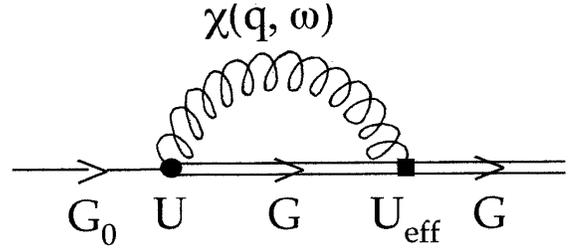,width=8cm,angle=0}}
\vspace{0.5ex}
\caption{Illustration of the coupling between holes or electrons and
spin fluctuations characterized by the susceptibility $\chi({\bf
q},\omega)$. This leads to Dyson's equation $G^{-1} = G_0^{-1} -
\Sigma$ describing the relation between the bare Green's function
$G_0$ and the renormalized one $G({\bf k},\omega)$. $U$ ($U_{\rm
eff}$) denotes an effective (renormalized) coupling constant.}
\label{diagram}
\end{figure}

{\bf  1.~Theoretical background:}
Similar to the electron-phonon case, the coupling between
the quasiparticles and the spin excitations should influence
characteristically the energies $\omega_{\bf k}$ of the hole- or
electron carriers
\begin{equation}
\omega_{\bf k} = \epsilon_{\bf k} +
\mbox{Re } \Sigma({\bf k},\omega) \quad .
\label{disp}
\end{equation}
Here, $\epsilon_{\bf k}$ refers to the bare energy dispersion of the
quasiparticles assuming no interaction with the spin excitations or
phonons, {\it i.e.} a tight-binding energy dispersion for the
CuO$_2$-plane.  In general, the self-energy $\Sigma ({\bf k},\omega)$
results from the coupling of the particles to spin excitations, see
Fig.  \ref{diagram} for an illustration.
Obviously, the self-energy is a functional of the spin susceptibility,
$\Sigma = \Sigma\{\chi\}$. In its simplest form, the latter is of
Ornstein-Zernicke form \cite{pines} that allows for a sharp enhancement
of fluctuations near the antiferromagnetic wave vector
{\bf Q}$=(\pi,\pi)$:
\begin{equation}
\chi({\bf q},\omega)=\frac{\chi_{\bf Q}}{1+
\xi^2({\bf q}-{\bf Q})^2 - 
i\frac{\omega}{\omega_{sf}}}.
\label{mmp}
\end{equation}
Here, $\chi_{\bf Q}$ is the value of the static spin susceptibility at
the wave vector {\bf Q}, $\xi$ is the magnetic correlation length, and
$\omega_{sf}$ is the characteristic frequency of spin fluctuations
(roughly the peak position in the imaginary part of Eq. (\ref{mmp})).
Due to the fact that $\chi({\bf q}, \omega)$ has only structures
around $\omega_{sf}$ and wave vector {\bf Q}, their influence on the
elementary excitations is expected to be very anisotropic at different
parts of the Brillouin Zone.

\begin{figure}[t]
\vspace{0cm}
\hspace*{0.3cm}
\centerline{\epsfig{file=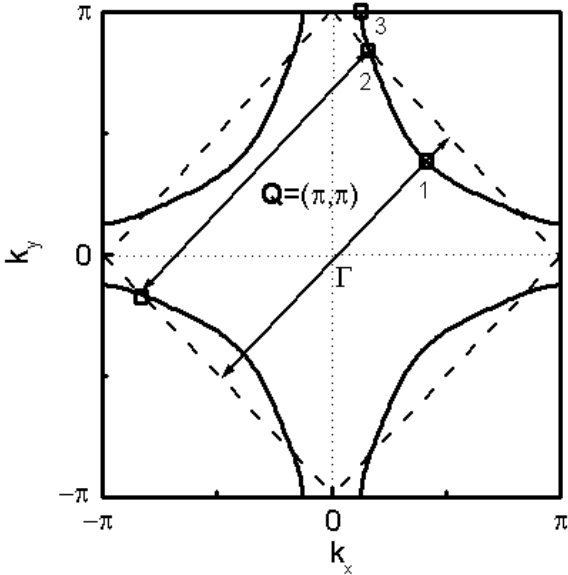,width=8cm,angle=0}}
\vspace{0.5ex}
\caption{Illustration of the anisotropy of the elementary
excitations. The solid line denotes the calculated Fermi surface for
the hole-doped cuprates in the first BZ using Eq.
(\ref{bareepsilon}). The dashed lines refer to the magnetic
Brillouin Zone that crosses the electronic Fermi surface close to
the $(\pi,0)$ points exactly at the 'hot spots' where the
antiferromagnetic wave vector ${\bf Q}=(\pi,\pi)$ connects two
pieces of the Fermi surface. At the ($\pi,0$) point and along the
diagonal the wave vector ${\bf Q}$ connects quasiparticle states
below the Fermi level only.  This allows us to define three
characteristic regions 1,2,3 at the Fermi level.}
\label{fermi}
\end{figure}

In order to illustrate the anisotropy of the elementary excitations we
show in Fig. \ref{fermi} the calculated Fermi surface for hole-doped
cuprates \cite{remark1}.  For the calculation we use the tight-binding
energy dispersion for the CuO$_2$-plane
\begin{equation}
\epsilon_{\bf k} = -2t\left( \cos k_x  + \cos k_y
\right) + 4t'\cos k_x \cos k_y -\mu
\quad .
\label{bareepsilon}
\end{equation}
Here, $t$ and $t'$ refer to the hopping of a hole (electron) between
nearest, next-nearest sites on the square lattice, and $\mu$ is the
chemical potential that defines the doping. In Eq. (\ref{bareepsilon})
and in the following we set the lattice constant to unity.  One
clearly sees that the scattering of quasiparticles by spin
fluctuations in cuprates is anisotropic.  First, the antiferromagnetic
wave vector ${\bf Q}=(\pi,\pi)$ connects exactly quasiparticles at the
Fermi level close to the ($\pi,0$) points of the first Brillouin Zone.
These quasiparticles
experience the strongest coupling to antiferromagnetic spin
fluctuations. Quasiparticles at the diagonals of the BZ are {\it not}
connected by the ${\bf Q}$ and thus have smaller scattering by spin
fluctuations at the Fermi level. The corresponding point on the Fermi
surface is called {\it cold spot} or {\it nodal point}.  Furthermore,
the quasiparticle states at the Fermi level close to the $M$ point in
the first BZ are usually called antinodal points.  Note that the
quasiparticle states at cold and hot spots which are connected by
${\bf Q}$ lie {\it close} {\it to} the Fermi level. This will be
important later for discussing the kink feature.  Therefore, we may
define three characteristic regions at the Fermi surface regarding
their sensitivity to coupling to antiferromagnetic spin fluctuations.

{\bf 2. Experimental findings:}
Let us now discuss the experimental situation {\it close} {\it to} the
Fermi level as measured by ARPES.  The combined study of energy
distribution curves (EDC) and momentum distribution curves (MDC)
allows the study of the quasiparticle excitations close and well below
the Fermi level up the energies of $200$ meV. Most importantly, recent
ARPES studies by various groups \cite{valla,johnson,shen,shen1} reveal
a kink structure in the energy dispersion at energies about $60
\pm 15$meV below the Fermi energy and along the nodal direction $(0,0) \to
(\pi,\pi)$ direction of the first BZ.  Furthermore, this kink
structure appears in the normal state and almost does not 
change if one enters the superconducting state.

\begin{figure}[t]
\vspace{0cm}
\hspace*{0.1cm}
\centerline{\epsfig{file=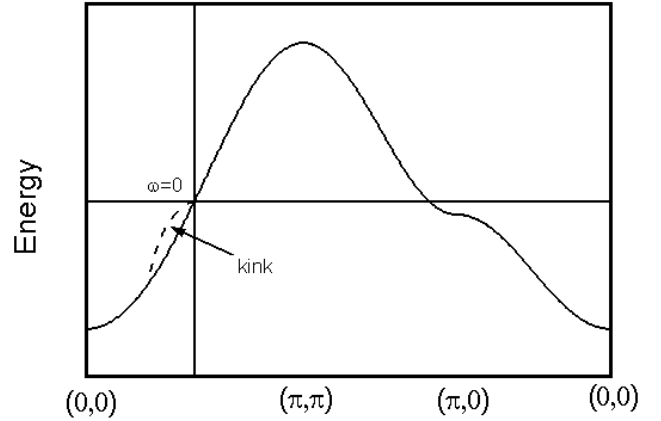,width=8.8cm,angle=0}}
\vspace{0.5ex}
\caption{Calculated bare tight-binding energy dispersion for the
hole-doped cuprates in the normal state using Eq. (\ref{bareepsilon}).
The dashed curve illustrates the changes due to renormalization,
$\omega_{\bf k} = \epsilon_{\bf k} + \mbox{Re}\Sigma ({\bf k},
\omega)$. Due to structure in Re$\Sigma$ the energy dispersion shows
a kink feature along the $(0,0) \to (\pi,\pi)$ direction in the
first Brillouin Zone.}
\label{disp1}
\end{figure}

We illustrate the formation of the kink in Fig. \ref{disp1} where we
show the unrenormalized tight-binding energy dispersion $\epsilon_{\bf
  k}$ in the normal state along the route $(0,0) \to (\pi,0) \to
(\pi,\pi) \to (0,0)$ of the first BZ.
Below E$_f$, the measured energy dispersion $\omega_{\bf k}$ along
$(0,0) \to (\pi,\pi)$ has changed compared to the bare tight-binding
case due to the self-energy corrections as it follows from Eq.
(\ref{disp}).  We define the kink at the inflection point of the
dashed curve where the renormalized dispersion tends to approach again
the bare dispersion $\epsilon_{\bf k}$.  Recent studies by Dessau and
co-workers \cite{dessau} reveal no kink structure along $(\pi,0) \to
(\pi,\pi)$ direction in the normal state, but only in the
superconducting state.

{\bf 3. Doping and ${\bf k}$-dependence of the kink:} Previously we
have shown\cite{manske} that from the momentum and energy conservation
one expects changes of the quasiparticle velocities $v_{\bf k} =
\partial\epsilon / \partial {\bf k}$ in the nodal direction for
$\omega \simeq \omega_{sf}$ and ${\bf k} - {\bf k'} = {\bf q} \simeq
{\bf Q}$.  This leads to the kink in the $(0,0) \to (\pi,\pi)$
direction that has been observed by angle-resolved photoemission
spectroscopy (ARPES) \cite{valla,johnson,shen,shen1}.  Here, we extend
our previous analysis and discuss in more detail the doping dependence
and the anisotropy of the kink feature as a function of (${\bf k} -
{\bf k}_F$) and energy $\omega_{\bf k}$ in the normal and
superconducting state, for different routes ${\bf k} - {\bf k}_F$ in
the Brillouin zone: $(0,0) \to (\pi,\pi)$, $(0,0) \to (0,\pi)$, and
$(\pi,0) \to (\pi,\pi)$. Due to the fact that the superconducting
order parameter has nodes along $(0,0) \to (\pi,\pi)$ and maxima for
$(0,0) \to (0,\pi)$ one expects additional changes in the kink
structure as observed recently \cite{kaminski,dessau}. This should
help to contrast renormalization due to spin fluctuations to what we
get due to electron-phonon interaction.  Furthermore, we investigate
the interdependence of the elementary excitations with the resonance
peak that is observed in hole-doped cuprates by inelastic neutron
scattering (INS) below $T_c$.  It follows also from Eq.  (\ref{disp})
and Im$\Sigma({\bf k},\omega) \propto \omega^2 \to \omega$ that the
temperature dependence of the kink feature should reflect
characteristically the coupling to spin fluctuations and be different
from the electron-phonon coupling case.

Note that in the case of electron-doped cuprates neither a resonance
peak nor a kink feature is present. We will also show that the kink
feature is not restricted to the cuprates, but is also expected for
other novel superconductors where quasiparticles couple strongly to
spin excitations. A particularly interesting case might be
$\mbox{Sr}_2\mbox{RuO}_4$ with large anisotropic behavior of
$\chi_{zz}({\bf q},\omega)$ and in-plane $\chi_{+-}({\bf q},\omega)$
\cite{ishida,eremin02}.  Generally, a strong nesting behavior of the
Fermi surface might yield pronounced kink features.

This paper is organized as follow: In Section \ref{theory} we present
the theory, and in Section \ref{res} we discuss our results for the
kink structure, its doping- and temperature dependence for hole- and
electron-doped cuprates. Finally, in Section \ref{summary} we
summarize our analysis and contrast renormalization due to spin
fluctuations and phonons.

\section{Theory}\label{theory}

{\bf A. One-band Hubbard model:}
The elementary excitations and the spin excitations are key quantities
determining the properties of the cuprates and other superconductors
with strong correlations and magnetic activity. The quasiparticles,
holes or electrons, interact strongly with spin fluctuations and also
with phonons. However, phonons and spin fluctuations differ with
respect to their doping dependence and anisotropy.  This is clearly
demonstrated by neutron scattering experiments, for example. Also the
different behavior of hole- and electron-doped cuprates and the
feedback of superconductivity on $\chi({\bf q},\omega)$ is important.

In this paper we employ an effective one-band Hamiltonian. This is
justified because upon hole doping antiferromagnetism disappears due
to Zhang-Rice singlet formation and quenching of Cu-spins.  In this
one-band picture the Coulomb interaction between the quasiparticles
refers to an effective interaction (i.e. the Hubbard $U$) within the
conduction band.  Then, further doping increases the carrier mobility
and a system of strongly correlated quasiparticles occurs. In the
overdoped case less magnetic activity is present yielding usual Fermi
liquid.  We assume $U \simeq W/2$ ($W=$ bandwidth) independent of
the doping concentration.

The main physics of a single CuO$_2$-plane is the two-dimensional
one-band Hubbard model given by
\begin{equation}
H = - \sum_{\langle ij \rangle \, \sigma}
t_{ij}\left( c_{i\sigma}^+ c_{j\sigma} +
c_{j\sigma}^+ c_{i\sigma}\right)
+ U\, \sum_i n_{i\uparrow}n_{i\downarrow}
\quad ,
\label{eq:hubbard}
\end{equation}
where $c_{i\sigma}^+$ creates an electron with spin $\sigma$ on site
$i$, $U$ denotes the on-site Coulomb interaction, and $t_{ij}$ is the
hopping integral. After diagonalization of the first term, one arrives
at the bare tight-binding energy dispersion given by Eq.
(\ref{bareepsilon}).  The description of the electron- and hole-doped
cuprates within a one-band approximation is possible if one takes into
account different parameters and quasiparticle dispersion\cite{king}.
Note that in the case of electron doping the electrons occupy the
copper $d$-band, while in the hole-doped case holes refer mainly to
the oxygen $p$-states yielding different dispersion parameters.
Furthermore, the energy dispersions for optimally hole-doped
La$_{2-x}$Sr$_x$CuO$_4$ (LSCO) and electron-doped
Nd$_{2-x}$Ce$_x$CuO$_4$ (NCCO) behave differently around $(\pi,0)$
point.  While in the case of LSCO the flat band (leading to the van
Hove singularity in the density of states) lies close to the Fermi
level, in NCCO the flat band is approximately $300$ meV below the
Fermi level. Then, using $t=250$meV and $t'=0.1t$, one describes the
hole-doped LSCO dispersion, whereas $t=138$meV and $t'=0.3t$ are
needed for the description of the electron-doped NCCO compound fitting
earlier photoemission data. These parameters including an intermediate
coupling $U=4t$, will be used for our calculations of various physical
quantities in the normal and superconducting state of electron- and
hole-doped cuprates \cite{remark2}.

{\bf B. Generalized Eliashberg equations:} In this one-band model, we
assume that the {\it same} electrons (holes) are participating in the
formation of antiferromagnetic fluctuations and in Cooper-pairing due
to the exchange of these fluctuations. Thus, both the magnetic
susceptibility and the quasiparticle self-energy must be calculated
self-consistently.  This is possible in the FLEX approximation
\cite{bickers,tew,ben,carb}. In this approach the dressed one-electron
Green's function are used to calculate the charge and spin
susceptibilities.  These susceptibilities are then used to construct
an effective Berk-Schrieffer-like \cite{berk} pairing interaction
$V_{\rm eff}$ describing the exchange of charge and spin fluctuations.
The generalized Eliashberg equations are derived in Appendix
\ref{appa}. In order to demonstrate the significant role of $V_{\rm
  eff}$ in our work, we also show in Fig.~\ref{veff} its corresponding
diagramatic representation.  Note, in general, if the Cooper-pairing
and the effective pairing potential $V_{\rm eff}$ are generated by the
same quasiparticles (solid lines in Fig. \ref{veff}), strong
self-energy and feedback effects on $G({\bf k},\omega)$ and $\chi({\bf
  q},\Omega)$ are expected \cite{migdal0}.

To be more precise, we write down the quasiparticle self-energy
components $\Sigma_{\nu}$ ($\nu=0,3,1$) with respect to the Pauli
matrices $\tau_{\nu}$ in the Nambu
representation\cite{nambu,schrieffer}, {\it i.e.}
$\Sigma_0=\omega(1-Z)$ (mass renormalization), $\Sigma_3 = \xi$
(energy shift), and $\Sigma_1 = \phi$ (gap parameter).
They are given by
\begin{eqnarray}
\Sigma_{\nu}({\bf k},\omega) & = & N^{-1}\sum_{\bf k'}
\int_{0}^{\infty} d\Omega \, V_{\rm eff} ({\bf k-k'},\Omega)
\nonumber\\[1ex]
& &\times \int_{-\infty}^{+\infty} d\omega'
I(\omega,\Omega,\omega') \, A_{\nu} ({\bf k'},\omega')
\label{flex1}
\end{eqnarray}
with
\begin{equation}
V_{\rm eff} =
\left[ P_s({\bf k}-{\bf k'},\Omega) - (\delta_{\nu 1} -
\delta_{\nu0} - \delta_{\nu 3}) \, P_c({\bf k-k'},\Omega)
\right] .
\end{equation}
This is a generalization of figure \ref{diagram}.  $P_s$ and $P_c$
denote the spectral density of the spin and charge excitations,
respectively, and are defined in Eqs. (\ref{ps}) and (\ref{pc}).  The
second part of Eq. (\ref{flex1}) is given in Eqs.  (\ref{flex2}) and
(\ref{flex3}).  It is interesting to remark that the above formulae
remain valid even if the elementary excitations and the magnetic
activity that controls $V_{\rm eff}$ would result from different
quasiparticles.

The generalized Eliashberg equations allow us to calculate all the
properties of the system self-consistently like superconducting phase
diagram, elementary excitations, the superconducting order parameter,
energy dispersion and dynamical spin susceptibility, for example
\cite{mabe,mabed}.  Note, in Eq.  (\ref{flex1}) self-energy effects
due to phonons are neglected. Their contribution will be discussed in
section \ref{summary} and in Appendix \ref{appb}.

\section{Results and Discussion}\label{res}

\subsection{ Anisotropic Renormalization}

{\bf 1. Nodal direction:} We start the discussion analyzing the
spectral density of hole-doped superconductors in the normal state.
The spectral density reveals the elementary excitations and in
particular the renormalized energy dispersion.  First, we present our
results for the spectral density along the nodal $(0,0) \to (\pi,\pi)$
direction in the first BZ.

\begin{figure}[t]
\centerline{\epsfig{file=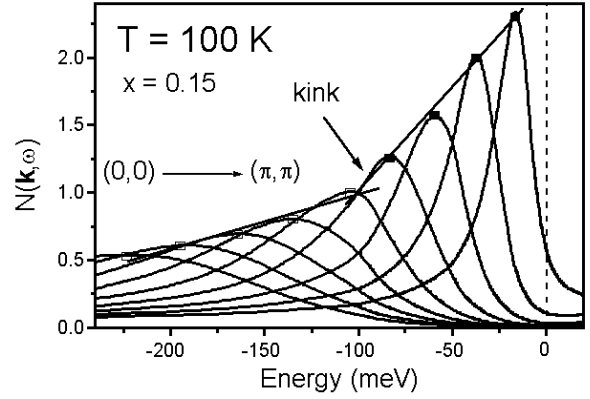,width=8cm,angle=0}}
\vspace{0.5ex}
\caption{Calculated spectral density $N({\bf k},\omega)$ in the
normal state along the nodal $(0,0) \to (\pi,\pi)$ direction (from
left to right) as a function of frequency in the first Brillouin
zone (BZ).  The peak positions (connected by the solid line to guide
the eye) refer to the renormalized energy dispersion $\omega_{\bf
k}$. One clearly sees the kink structure at an energy
approximately $\omega_{kink} = 75 \pm 15$meV that results from
coupling of the quasiparticles to spin fluctuations.  }
\label{nodal}
\end{figure}

In Fig. \ref{nodal} we show the calculated spectral density $N({\bf
  k},\omega)$, i.e. the local density of states, as a function of
frequency and momentum ${\bf k}- {\bf k}_F$. The peak positions
correspond to the renormalized energy dispersion.  Due to coupling of
holes to antiferromagnetic spin fluctuations the quasiparticle
dispersion changes its slope and shows a pronounced kink feature at
the energy $\omega_{kink} \approx 75 \pm 15$meV.

How can one understand the kink feature in a simple way?  At the first
glance the occurrence of a kink in the nodal direction seems to be
surprising, since the main interaction of the carriers with spin
fluctuations occurs at the hot spots while the kink feature is present
along the diagonal of the BZ close to the cold spots.  This argument,
however, considers only the quasiparticles exactly at the Fermi level.
Away from the Fermi level but close to it (along $(0,0) \to (\pi,\pi)$)
the quasiparticles couple strongly to spin fluctuations, as can be seen from
Fig. \ref{fermi}. Most importantly, as follows from Fig.
\ref{fermi}, the largest scattering will occur at values of ${\bf k} -
{\bf k}_F = {\bf Q}$ and $\omega = \omega_{sf}$. To be more precise,
let us rewrite Eq. (\ref{flex1}) as
\begin{figure}[t]
\vspace{-1cm}
\centerline{\epsfig{file=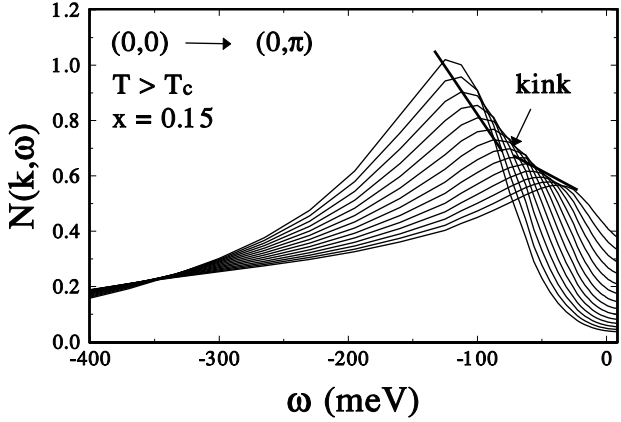,width=7cm,angle=-90}}
\vspace{0.5ex}
\caption{Spectral density $N({\bf k},\omega)$ as a
function of frequency along the $(0,0) \to (0,\pi)$ direction of the
first BZ in the normal state calculated from the generalized
Eliashberg equations. Again, the peak positions reveal the
renormalized energy dispersion $\omega_{\bf k}$.  A kink occurs at
similar energy as in the nodal direction. Because of inelastic
scattering of holes on spin fluctuations close to $(0,\pi)$, $N({\bf
k},\omega)$ becomes also broader. Note that, in contrast to the
nodal direction, one does not cross the Fermi level in the $(0,0)
\to (0,\pi)$ direction. Instead, one reaches the flat part of the
tight-binding band.}
\label{unphys}
\end{figure}
\begin{eqnarray}
\Sigma({\bf k},i\omega_n)
& = &
- T^2 \sum_{\omega_m, \nu_m } 
\sum_{\bf k',q} \tilde{\tau}_0 
G({\bf k}-{\bf k'},i\omega_n - i\nu_m) \tilde{\tau}_0 U^2
\nonumber\\
& \hspace*{-1cm}\times &
\hspace*{-0.5cm}\frac{1}{2} \mbox{Tr}\left[\tilde{\tau}_0 
G({\bf k}+{\bf q},i\omega_m + i\nu_m) \tilde{\tau}_0 
G({\bf q},i\omega_m)\right]
\label{self}
\end{eqnarray}
and approximate the Green's function by its non-interacting part
\begin{equation}
G({\bf k},i\omega_n) \approx G_0({\bf k},i\omega_m) = 
\frac{i\omega_n \tilde{\tau}_0 +\epsilon_{\bf k} \tilde{\tau}_3 -
\phi_{\bf k} \tilde{\tau}_1 }{(i\omega_n)^2 - E_{\bf k}^2}
\quad.
\end{equation}
with $E_{\bf k}^2= \epsilon_{\bf k}^2 + \phi_{\bf k}^2$. Thus, after
little algebra one obtains on the real axis \cite{brinkman}
\begin{eqnarray}
\Sigma({\bf k},\omega) \approx -\frac{U^2}{4}
& &
\sum_{\bf k'} \int_0^{\infty} 
d\omega' \frac{\mbox{Im} \chi_{\mbox{RPA}}(
{\bf k-k'},\omega')}{\omega-\omega'-
E_{\bf k'}} 
\nonumber \\[1ex]
& \hspace*{-1.3cm}\times &
\hspace*{-0.5cm}\left[\coth\left(\frac{\omega'}{2T}\right)
- \tanh \left(\frac{\omega'-\omega}{2T}\right)\right] 
\quad.
\label{eqimsigma}
\end{eqnarray}
The imaginary part of the spin susceptibility may be obtained within
the RPA and is approximately given by Ornstein-Zernicke expression
(see Eq.(\ref{mmp})).  This self-energy $\Sigma$ now enters Eq.
(\ref{disp}). It is important that the self-energy is mainly
frequency-dependent, while the bare dispersion of the carriers is not.
Already in the normal state, $\Sigma({\bf k},\omega)$ has a maximum
reflecting a corresponding maximum of $\mbox{Im }\chi$ at ${\bf q}
\approx {\bf Q}$ and $\omega' \approx \omega_{sf}$.  Note that
$\omega_{sf}$ can be determined according to Moriya and Ueda and
paramagnon theory from $\chi_{RPA}=\chi_0/(1 - U \chi_0)$ or
equivalentlt from $\chi_0^{-1}({\bf q}\simeq {\bf Q},
\omega\simeq\omega_{sf}) - U = 0$ \cite{remark3}.  Obviously,
$\omega_{sf}$ is strongly doping dependent.  This will be discussed
later.  Then, the kink position follows from the pole of the
denominator of Eq.  (\ref{eqimsigma}). This leads to the 'kink
condition'
\begin{equation}
\omega_{kink} \approx E_{\bf k-Q} + \omega_{sf}(x)
\quad.
\label{nodalsf}
\end{equation}
This gives an estimate of the position of the
kink.  Furthermore, since the superconducting gap is zero for $\omega
=0 $, but not for $\omega = \omega_{sf}$, the kink feature along the
nodal direction $(0,0) \to (\pi,\pi)$ will change only slightly below
$T_c$. This we have demonstrated previously \cite{manske}.

{\bf 2.~${\bf (0,0) \longrightarrow (0,\pi)}$ direction:} In order to
see whether the kink feature is present in other directions of the
Brillouin Zone, we show in Fig. \ref{unphys} the evolution of the
spectral density along the $(0,0) \to (0,\pi)$ direction. Despite of
the fact that along this direction we do not cross the Fermi level,
the kink feature is still present and is found at an energy similar to
the one for the nodal $(0,0) \to (\pi,\pi)$ direction.  This indicates
that the occurrence of the kink feature is not connected to some
specific conditions which might be present only along the $(0,0) \to
(\pi,\pi)$ direction.  Instead, the kink is characteristic for all
direction where ${\bf k} - {\bf k}_F\simeq {\bf Q}$ and $\omega \simeq
\omega_{sf}$. Also below $T_c$ we find that the kink feature is
present in the $(0,0) \to (0,\pi)$ direction (not shown).  Note, that
our results are in fair agreement with experimental data \cite{shen}.

\begin{figure}[t]
\vspace*{-1cm}
\centerline{\epsfig{file=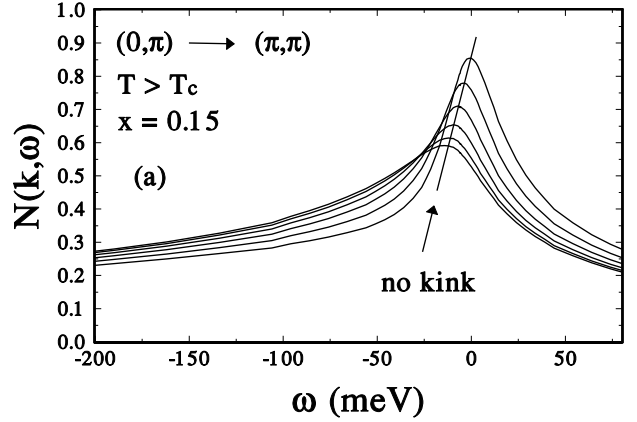,width=7cm,angle=-90}}
\vspace*{-0.1cm}
\centerline{\epsfig{file=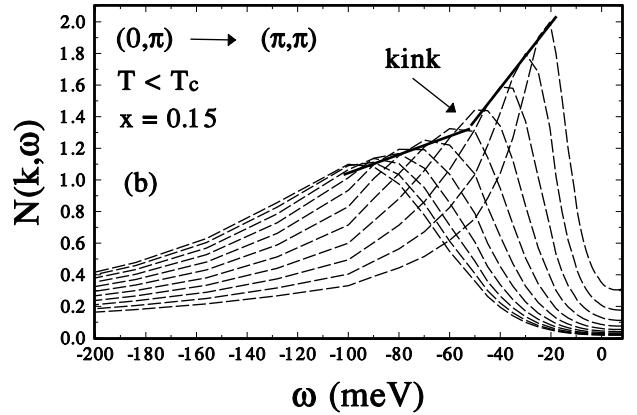,width=7cm,angle=-90}}
\vspace{0.5ex}
\caption{Calculated spectral density
$N({\bf k},\omega)$ along the antinodal $(\pi,0) \to (\pi,\pi)$
direction as a function of frequency in the first BZ in the normal
(a) and superconducting (b) state. Due to the flat band close to the
Fermi level the spectral density shows no kink structure in the
normal state. Below T$_c$ the superconducting gap
$\phi(\omega)$ opens yielding a kink structure in the spectral
density that occurs at the energies $\omega_{kink} \approx 50 \pm
10$ meV at optimal doping.}
\label{gammaM}
\end{figure}

{\bf 3.~Antinodal direction:}
In Fig. \ref{gammaM}(a) we show our results for $N({\bf k},\omega)$
along the $(\pi,0) \to (\pi,\pi)$ route, i.e. the antinodal direction,
of the first BZ in the normal state.  Note, that the spectral density
at the $(0,\pi)$ point is broader than at the antinodal point due to
stronger coupling to spin excitations peaked at ${\bf q} = {\bf Q} =
(\pi,\pi)$ as discussed in Fig. \ref{fermi}.  Clearly, no kink is
present.  The absence of a kink structure can be explained with the
flat structure of the $\mbox{CuO}_2$-plane around the $M$ point (see
Fig.  \ref{disp1}). Simply speaking, for a flat band the frequency
dependence of $\Sigma$ in Eq. (\ref{disp}) does not play a significant
role and therefore no change of the velocity and no kink structure is
present.

What does happen in the superconducting state?  Below $T_c$ the
superconducting gap $\phi({\bf k},\omega)$ opens rapidly for
decreasing temperature $T$ and becomes maximal in momentum space
around the $M$ point reflecting the momentum dependence of the
effective pairing interaction (see Eq. (\ref{flex1})). In addition,
due to the frequency dependence of the gap the flat band around $M$
disappears.

In Fig. \ref{gammaM}(b) we show results for $N({\bf k},\omega)$ at a
temperature $T = 0.5T_c$ where the superconducting gap has opened. A
kink structure around $\omega_{kink} \approx 50 \pm 10$ meV is present
reflecting the magnitude of $\phi$. Hence, in the $(\pi,0) \to
(\pi,\pi)$ direction this kink feature is only present below $T_c$ and
connected to the feedback effect of $\phi$ on the elementary
excitations. We will show later that this feedback is also important
for the resonance peak seen in INS.  

Note, that the superconducting gap
$\phi({\bf k},\omega)$ is calculated self-consistently in our theory
and reflecting the underlying spin fluctuations which dominate the
pairing potential $V_{\rm eff}$. Therefore, the occurrence of a kink
structure {\it only} below $T_c$ in the antinodal direction is a
direct fingerprint of the spin excitation spectrum \cite{prb2001}.
Furthermore, as we will discuss below, $\mbox{Im }\chi ({\bf
  Q},\omega)$ entering in Eq. (\ref{eqimsigma}) is peaked at the
resonance frequency $\omega_{res}$ (roughly at $\omega_{sf} +
\Delta$). Therefore, the kink condition is given by
\begin{equation}
\omega_{kink}
\approx E_{\bf k-Q} + \omega_{res}(x)
\quad.
\end{equation}

\begin{figure}[tt]
\vspace{-1cm}
\centerline{\epsfig{file=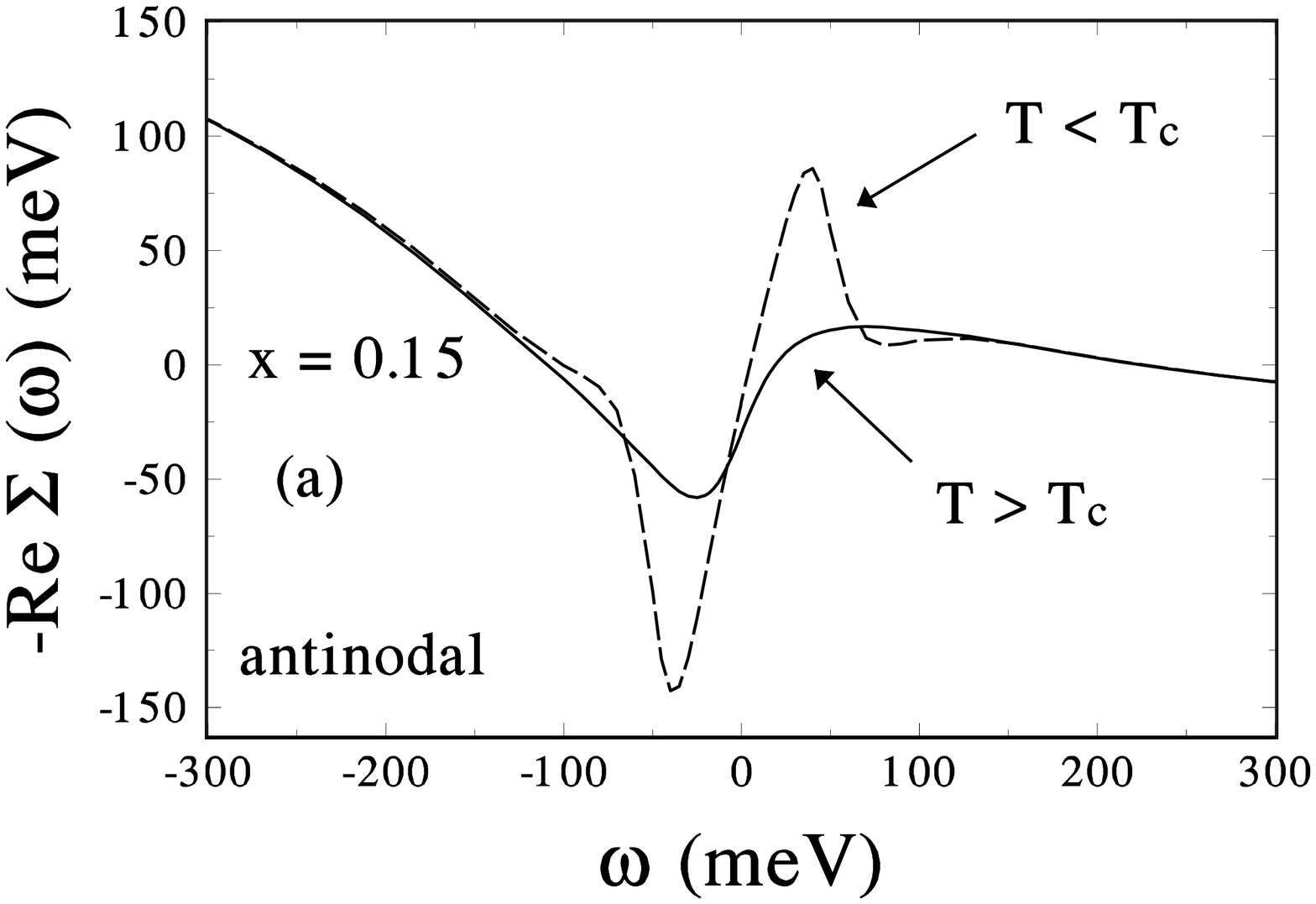,width=6.5cm,angle=-90}}
\vspace{-0.5cm}
\centerline{\epsfig{file=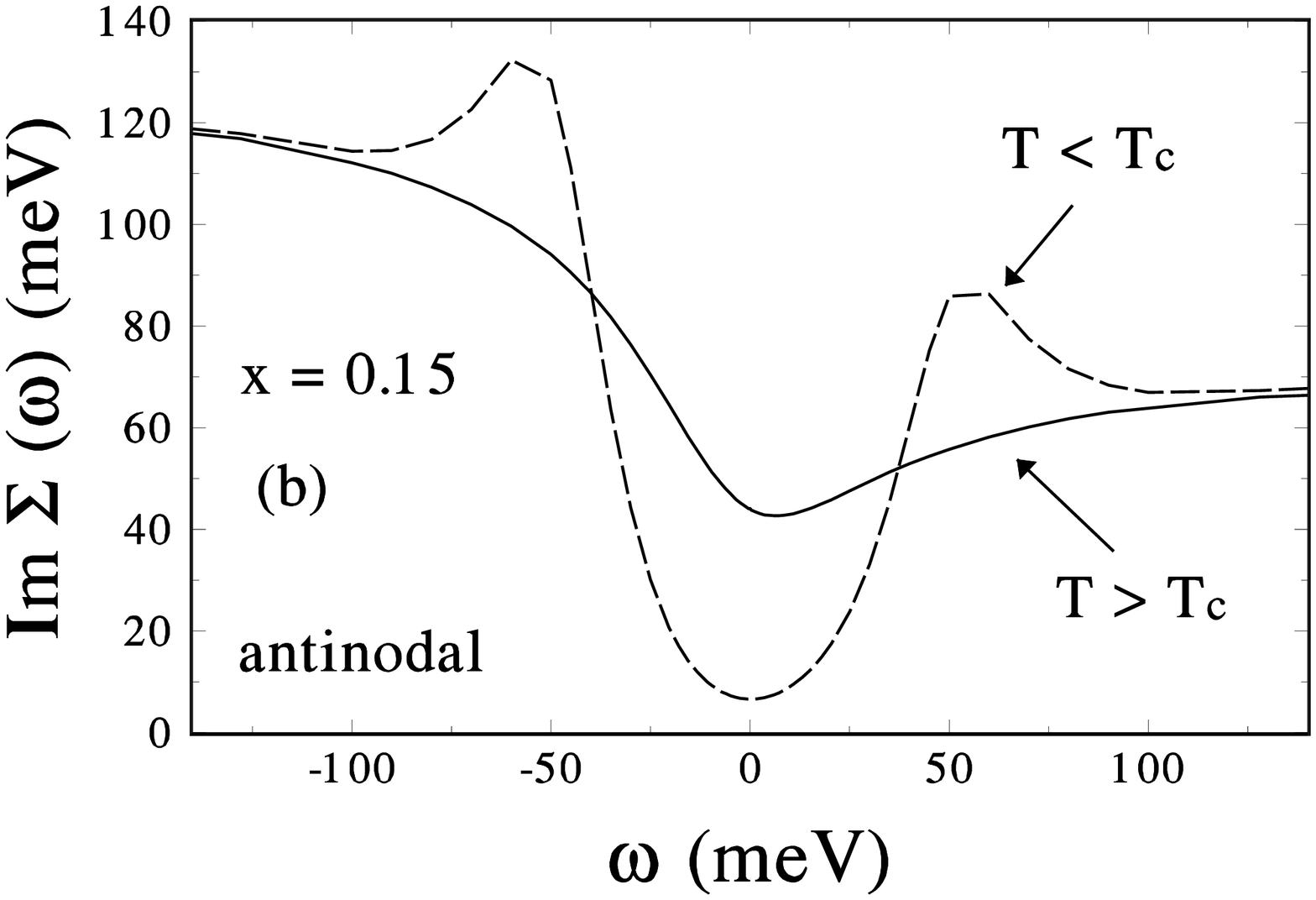,width=6.5cm,angle=-90}}
\vspace{0.5ex}
\caption{(a) Calculated frequency dependence of 
$\mbox{Re }\Sigma({\bf k}_a,\omega)$ at the antinodal point ${\bf
k}_a$ of the first BZ in the normal (solid curve) and
superconducting state (dashed curve). Due to the feedback effect of
the superconducting gap $\phi(\omega)$, a peak (dip) occurs for
$\omega > 0$ ($\omega < 0$) which roughly defines the position of
the kink structure. (b) The corresponding imaginary part at the
antinodal point $\mbox{Im }\Sigma({\bf k}={\bf k}_a,\omega)$ is
shown. Again, due to the feedback effect of $\phi(\omega)$, a
maximum occurs below $T_c$.  Note, that both $\mbox{Re }\Sigma$ and
$\mbox{Im }\Sigma$ are not fully antisymmetric (symmetric) with
respect to $\omega$ at optimum doping $x=0.15$.}
\label{sigre}
\end{figure}

In Fig. \ref{sigre}(a)
the frequency dependence of $\mbox{Re }\Sigma({\bf k}_a, \omega)$ in
the normal and superconducting state at the antinodal point ${\bf k} =
{\bf k}_a$ is shown. Due to the occurrence of the resonance feature in
$\mbox{Im }\chi({\bf Q},\omega)$ and the related feedback of the
superconducting gap $\phi(\omega)$, $\mbox{Re }\Sigma$ shows a
pronounced structure below $T_c$ at energies of about $\omega_{res} +
\Delta_0$.  Also the corresponding imaginary part, $\mbox{Im
  }\Sigma({\bf k}={\bf k}_a,\omega)$, shows a peak below $T_c$ (see
Fig. \ref{sigre}(b)).  This pronounced behavior is responsible for the
kink formation along $(\pi,0) \to (\pi,\pi)$ direction in the BZ.
Therefore, while the kink features are present along $(0,0) \to
(\pi,\pi)$ and $(\pi,0) \to (\pi,\pi)$ directions in the
superconducting state of hole-doped cuprates, their nature is
qualitatively different. Along the nodal direction the superconducting
gap is zero (for $\omega=0$) and thus the feedback effect of
superconductivity on the elementary and spin excitations is small.
Therefore, $\omega_{sf}$ determines mainly the formation of the kink
feature. On the other hand, along the antinodal direction the gap is
maximal and yields a strong feedback of superconductivity on $\chi$.
Thus, in the superconducting state $\omega_{res}$ and $\Delta_0$ yield
the kink structure along $(\pi,0) \to (\pi,\pi)$ direction that is not
present in the normal state.

\subsection{Doping dependence of renormalization}

The different reasons for the kink structures in hole-doped cuprates
along different directions in the first BZ will be also reflected in
their doping dependence. So far, the results we have shown were for
optimal doping concentration $x=0.15$ that refers to a band filling of
$n=0.85$\cite{mabed0}. Note, the superconducting transition
temperature $T_c$ behaves differently in the overdoped (OD) and
underdoped (UD) regime:
\begin{eqnarray*}
T_c & \propto & \Delta (T \to 0), \, \mbox{OD}
\\
T_c & \propto &  n_s (T \to 0), \, \mbox{UD}
\end{eqnarray*}
where $n_s$ is the superfluid density that is calculated
self-consistently from the generalized Eliashberg equations
\cite{mabed}.

In the antinodal $(0,\pi) \to (\pi,\pi)$ direction the kink is
only present below $T_c$ due to the feedback of $\phi(\omega)$. In
the OD case, $\phi(\omega)$ decreases reflecting a mean-field-like
behavior. Thus, the energy where the kink occurs must decrease with
overdoping:
\begin{equation}
\omega_{kink} (x) \propto \Delta_0 (x)
\quad .
\end{equation}
This behavior is indeed observed by Dessau and co-workers
\cite{dessau}. Note, the above argument remains true also in the
strongly OD case where no resonance peak in $\mbox{Im }\chi({\bf
  Q},\omega)$ occurs because the feedback effect of $\phi(\omega)$
should always be present.

\begin{figure}[t]
\vspace{-0.3cm}
\hspace*{0.3cm}
\centerline{\epsfig{file=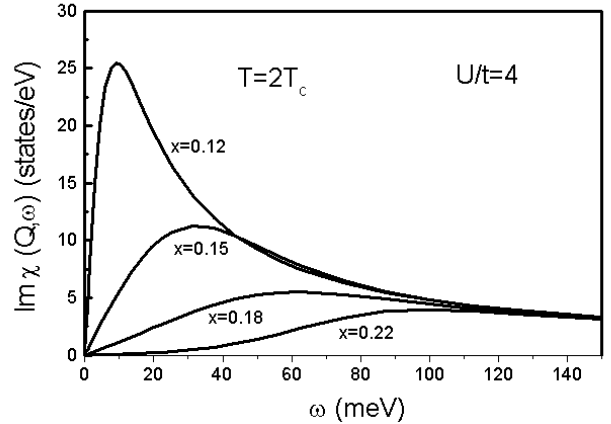,width=9cm,angle=0}}
\vspace{0.5ex}
\caption{Calculated paramagnon spectrum, i.e. the
dynamical spin susceptibility $\mbox{Im }\chi({\bf Q},\omega)$ at a
temperature $T=2T_c$ for different doping concentrations, $x=0.12$
(underdoped), $x=0.15$ (optimal doping), and $x=0.18$, $x=0.22$
(overdoped).}
\label{chi}
\end{figure}

Regarding the kink along the nodal $(0,0) \to (\pi,\pi)$ direction we
note the following. In Fig. \ref{chi} we show the calculated doping
dependence of \mbox{Im }$\chi$ at the antiferromagnetic wave vector
${\bf Q}$ versus frequency in the normal state. One clearly sees the
characteristic Ornstein-Zernicke behavior (see Eq. (\ref{mmp})) of
$\mbox{Im }\chi$,
\begin{equation}
\mbox{Im }\chi({\bf q}={\bf Q},\omega) \propto
\frac{\omega \, \omega_{sf}}{\omega^2 + \omega_{sf}^2}
\quad ,
\label{oz}
\end{equation}
and that $\omega_{sf}$ increases with increasing doping from
underdoped to overdoped cuprates. Since $\omega_{sf}$ determines the
kink position along $(0,0) \to (\pi,\pi)$ direction we expect
\begin{equation}
\omega_{kink}(x) \propto \omega_{sf} (x).
\quad .
\end{equation}
This is in qualitative agreement with experimental data \cite{johnson}
(for underdoped regime and optimally doped superconductors).  On the other
hand, the spectral weight of $\mbox{Im }\chi({\bf Q},\omega)$
decreases drastically with overdoping.  Therefore, the coupling of the
quasiparticles to spin fluctuations is getting much weaker in the OD
case. These two competing effects seem responsible for the
non-monotonic and weak doping dependence of the kink position in the
nodal direction \cite{dessau}.
\begin{figure}[t]
\hspace*{-0.3cm}
\centerline{\epsfig{file=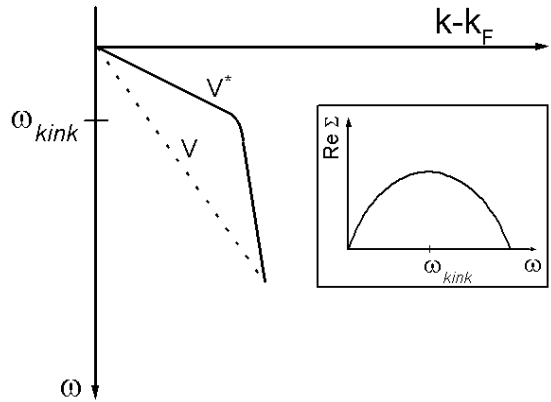,width=7.3cm,angle=0}}
\vspace{1ex}
\caption{Illustration of the kink position. We also show
$\mbox{Re }\Sigma$, since this controls the kink position and
$v^{*}$. The dashed curve refers to the bare dispersion. Note that
$v^{*} \to v$ for $\omega > \omega_{kink}$ reflects mainly the
width of the peak in $\mbox{Im }\chi$.}
\label{newillu}
\end{figure}

In Fig. \ref{newillu} we illustrate the kink feature resulting from
the renormalization ($\frac{d\Sigma'}{d\omega} \sim v^{*} \sim
(1+\lambda)$, $v \approx v_F$) of the bare dispersion. We estimate $2
\leq \lambda \leq 3$.  This renormalization is doping dependent and
stronger for underdoped hole-doped cuprates. Of course, we expect that
the position of the kink as well as the change of the quasiparticle
velocity ($v \to v^{*}$) are important fingerprints of the coupling to
spin fluctuations. Note, $v^{*} \to v$ for frequencies $\omega >
\omega_{kink}$ reflects mainly the width of the peak in $\mbox{Im
  }\chi$. Important is the slope ratio $v^{*}/v$ for $\omega <
\omega_{kink}$.

Another important behavior concerns the asymmetry between hole and
electron-doped cuprates.  Note that no kink feature has been reported
in the electron-doped cuprates \cite{nagaosa}. It is believed that the
electron-phonon coupling is much more pronounced in electron-doped
cuprates than in hole-doped ones. This is indicated, for example, by
the behavior of the resistivity $\rho \propto T^2$ in the normal state
at optimum doping and by the transition between $d_{x^2-y^2}$-wave
symmetry of the superconducting gap towards anisotropic $s$-wave as it
has been observed in several experiments \cite{skinta}.  Simply
speaking, the spin fluctuations in electron-doped cuprates are weaker
than in the hole-doped ones yielding a smaller $T_c$ and a smaller
superconducting gap \cite{prb2000}. Thus, no kink is present in the
nodal direction and also no kink occurs in the $(0,\pi) \to (\pi,\pi)$
direction below $T_c$. This is related to the fact that the flat band
around $(0,\pi)$ lies in electron-doped cuprates well below the Fermi
level and, therefore, it cannot be softened due to $\phi(\omega)$.

\subsection{Relation of kink and resonance peak}

\begin{figure}[t]
\vspace{0.5ex}
\centerline{\epsfig{file=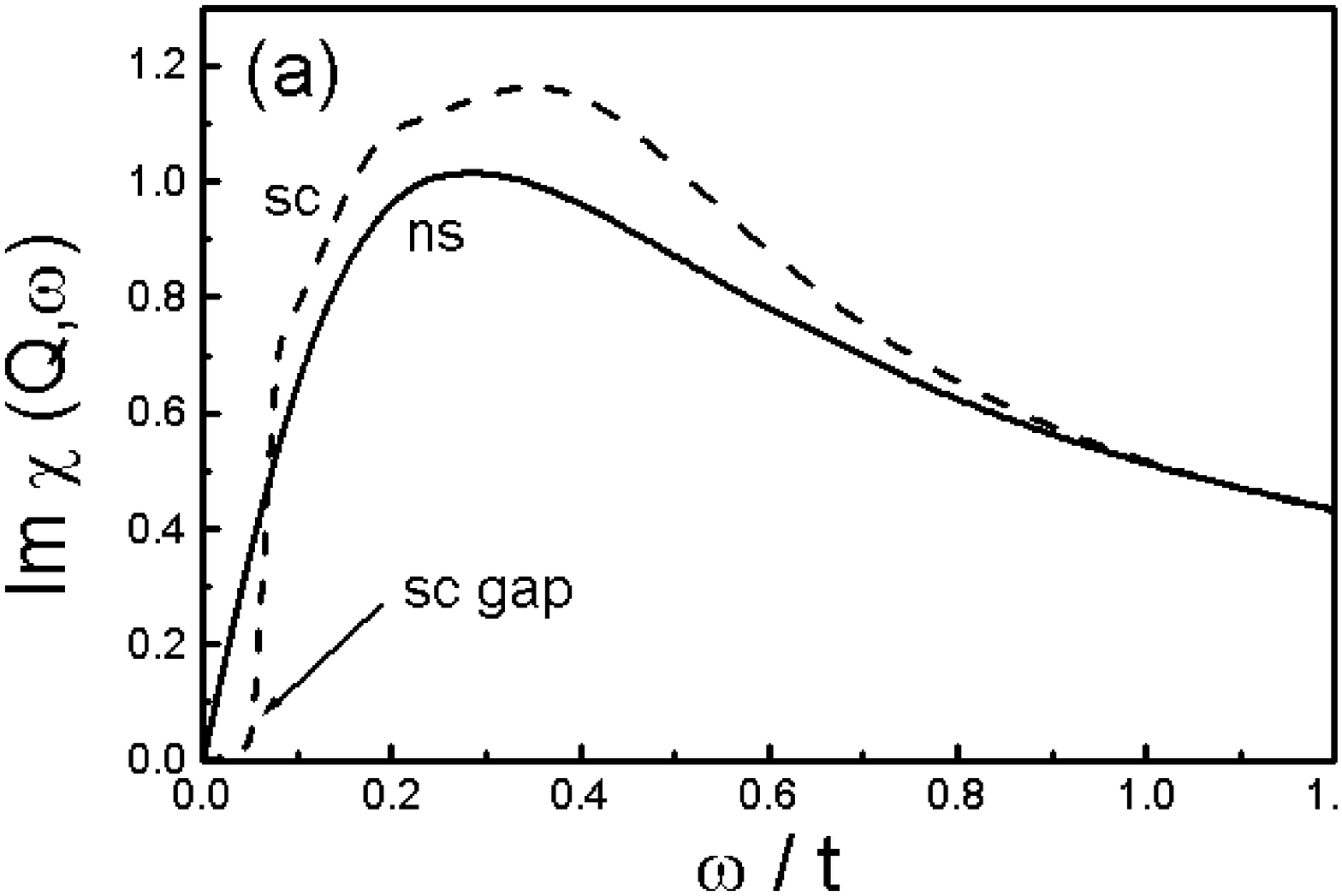,width=7.5cm,angle=0}}
\centerline{\epsfig{file=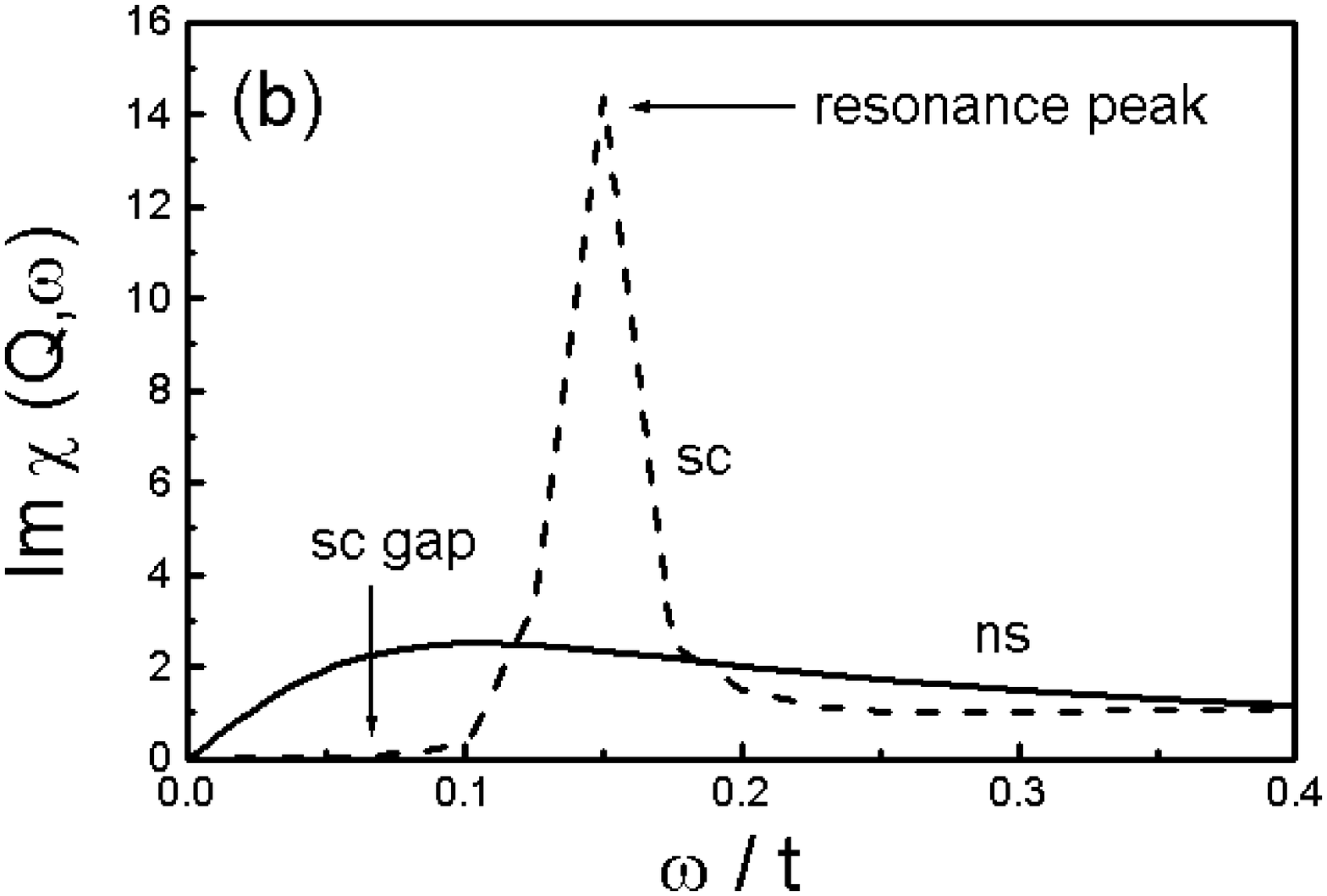,width=7.5cm,angle=0}}
\vspace{0.5ex}
\caption{Calculated feedback of superconductivity on the 
spin susceptibility $\mbox{Im }\chi({\bf q},\omega)$ for the
electron-doped (a) and hole-doped (b) cuprates at optimal doping
(x=0.15). The solid curves refer to the normal state (T=1.5T$_c$),
while the dashed curves denote the renormalized spin susceptibility
in the superconducting state at T=0.7T$_c$. Due to large
$\omega_{sf}=0.3t$ and the small superconducting gap, the feedback
of superconductivity is small in electron-doped cuprates.  Contrary,
due to a small $\omega_{sf}=0.09t$ in the hole-doped cuprates the
feedback of superconductivity fulfills a resonance condition for Im
$\chi$ yielding a strong renormalization of the spin excitation
spectrum and to a formation of the resonance peak. Note that the
hopping integral $t$ is different for hole- and electron-doped
cuprates as discussed in the introduction.}
\label{chielho}
\end{figure}

In Fig.~\ref{chielho} we show results
for the spin susceptibility $\mbox{Im }\chi_{RPA}({\bf Q},\omega)$ in
the optimally electron(a)- and hole(b)-doped cuprates in the normal
and superconducting state taking into account different tight-binding
energy dispersions \cite{king,prb2000}.
While in the normal state of hole-doped cuprates $\omega_{sf}$ is of
order of $25$ meV, in the electron-doped ones its value is much larger
($\omega_{sf} \approx 70$ meV) and $\mbox{Im }\chi$ is much less
pronounced. Therefore, antiferromagnetic spin fluctuations are much
weaker in the electron-doped cuprates due to weaker nesting of the
Fermi surface and less density of states due to the flat band well
below the Fermi level.  Regarding the superconducting state note, that
in the hole-doped cuprates a strong renormalization of the spin
fluctuation spectra occurs due to the feedback effect of
superconductivity and that $\Delta_0 \sim \omega_{sf}$ leading to a
resonance peak at $\omega=\omega_{res}$ (see Fig. \ref{chielho}(b)).
To be more precise, a resonance condition
\begin{equation}
\frac{1}{U_{cr}} = \mbox{Re }\chi_0({\bf q}={\bf Q},
\omega = \omega_{res})
\quad ,
\label{ucr}
\end{equation}
which signals the occurrence of a spin-density-wave collective mode,
must be fulfilled in order to observe a resonance peak \cite{prb2001}.
In electron-doped cuprates, the spin excitations do not obey Eq.
(\ref{ucr}) and thus only a rearrangement of spectral weight 
occurs below $T_c$, but no
resonance peak. Therefore, the kink feature is
intimately connected with the resonance peak. As we
see from Fig. \ref{chielho}(a) there is only a small feedback of
superconductivity below $T_c$ on $\mbox{Im }\chi$ in the
electron-doped cuprates due to $\omega_{sf} >> \Delta_0$. Thus, we
find also no kink feature in the superconducting state of
electron-doped cuprates in the antinodal direction.

\subsection{Anisotropic scattering rates}

\begin{figure}[t]
\vspace{-0.1cm}
%
\centerline{\epsfig{file=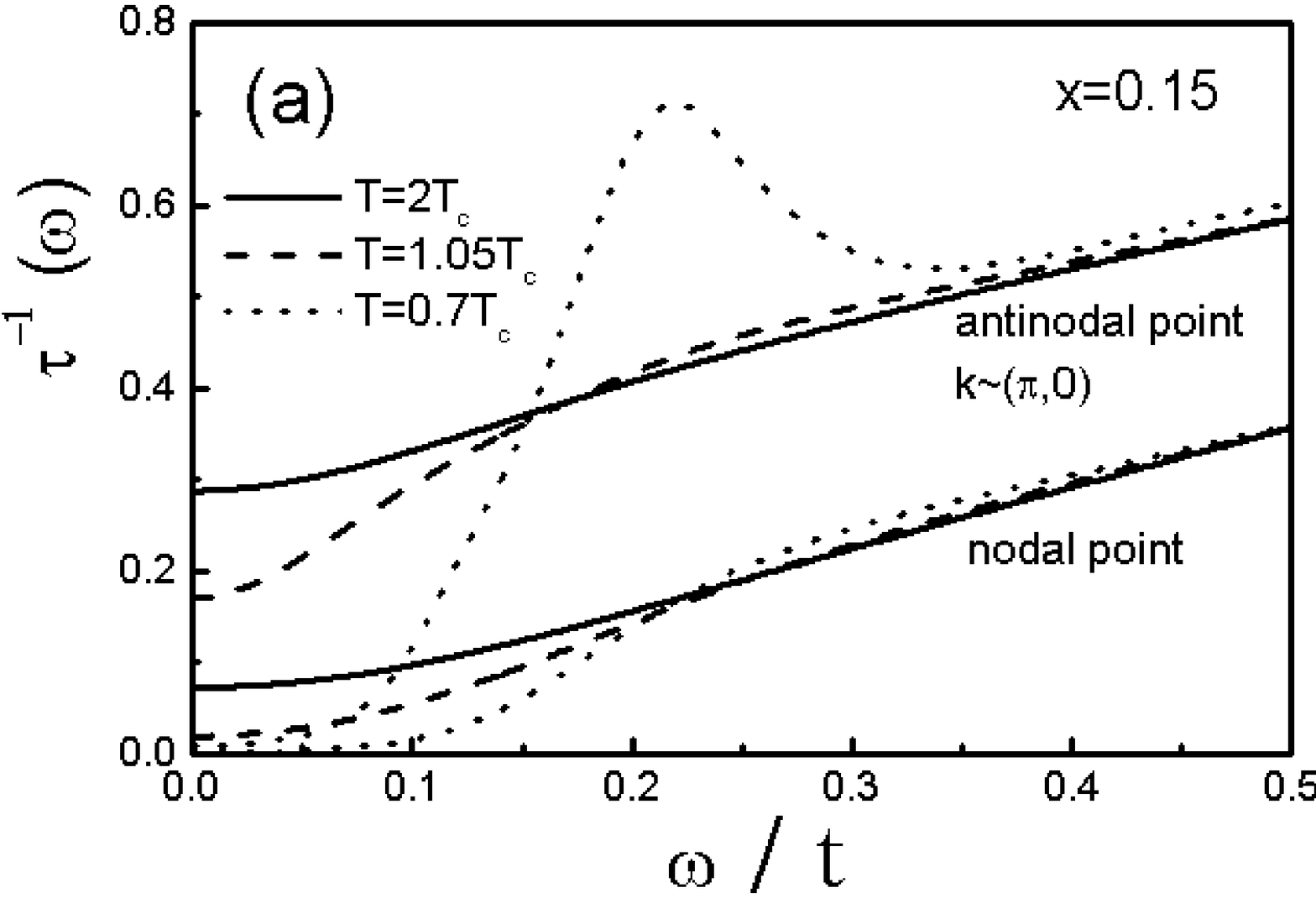,width=8cm,angle=0}}
\centerline{\epsfig{file=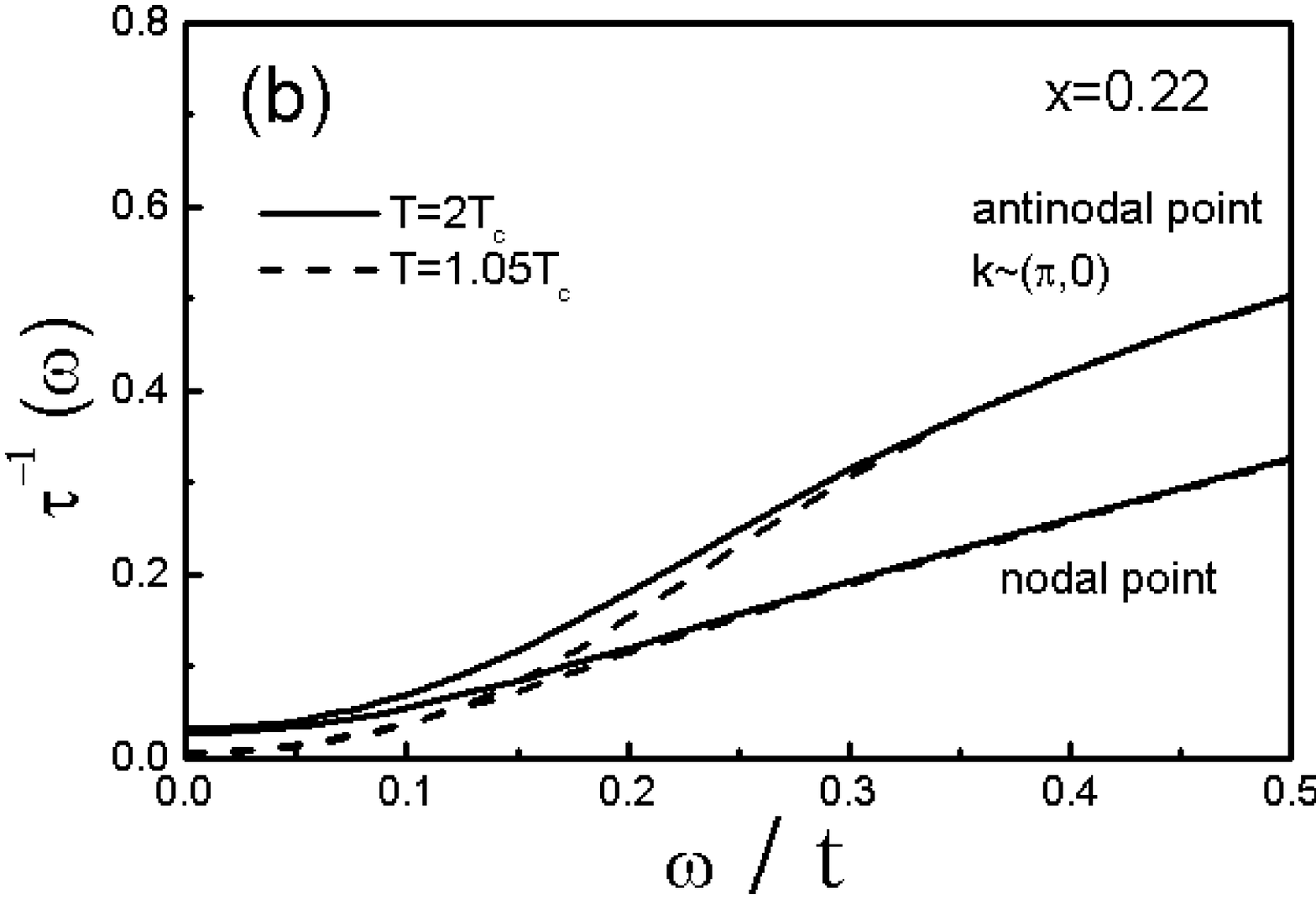,width=8cm,angle=0}}
\vspace{0.5ex}
\caption{Scattering rate $\tau^{-1}(\omega)$ of optimally (a) and
overdoped (b) hole-doped cuprates versus frequency at the nodal and
antinodal point of the Brillouin Zone (BZ) calculated at various
temperatures.  The anisotropy results from coupling to spin
fluctuations and is disappearing in the overdoped case. Thus, a
crossover from a non-Fermi liquid to a Fermi liquid behavior occurs.
Note also the feedback effect of superconductivity for different
parts of the BZ at optimal doping.}
\label{tau}
\end{figure}

Finally we discuss the anisotropy of the scattering rate
$\tau^{-1}(\omega)$ of hole-doped cuprates at different points on the
Fermi surface. In Fig. \ref{tau} we show our results for
$\tau^{-1}(\omega)$ at the antinodal point and the nodal point,
respectively, for optimal doping (a) and for the overdoped case (b)
for various temperatures. In Fig. \ref{tau}(a) one clearly sees that
the scattering rate is very anisotropic on the Fermi surface
reflecting the anisotropy of the coupling of elementary excitations to
spin fluctuations. In particular, $\tau^{-1}(\omega)$ in the normal
state is almost three times larger at the antinodal point than at the
nodal point. This agrees with recent ARPES experiments \cite{fink}.
Furthermore, we find that $\mbox{Im }\Sigma \propto \omega$
demonstrating a non-Fermi liquid behavior in the hole-doped cuprates.
Below $T_c$ at the antinodal point $\tau^{-1}(\omega)$ reveals a
strong feedback of superconductivity at energies $\omega_{res} +
\Delta_0$. At the nodal point the effect of superconductivity is
rather weak. In the overdoped cuprates the anisotropy between nodal
and antinodal points is strongly reduced and for $\omega \to 0$ almost
disappeared.  Most importantly the system then behaves more
Fermi-liquid-like. The latter is seen from Fig.  \ref{tau}(b) where
one observes a crossover from the $\mbox{Im }\Sigma \propto \omega$ to
the $\mbox{Im }\Sigma \propto \omega^2$ behavior.  This is also in
agreement with experimental observation \cite{johnson2}.

\section{Summary}\label{summary}

In summary, we have analyzed the elementary excitations in hole- and
electron-doped cuprates and the fingerprints of spin fluctuations 
on them.  The quasiparticles
around the antinodal points of the BZ experience the strongest
scattering on spin fluctuations yielding a non-Fermi liquid behavior.
In agreement with experimental data, we find that coupling of holes to
spin fluctuations yields a kink feature in the renormalized energy
dispersion.

{\bf Possible phonon contribution to the kink feature} One of the
interpretation of the kink structure in hole-doped cuprates has been
the electron-phonon interaction suggested by Lanzara {\it et al.}
\cite{shen1} Indeed, it is clear that phonons would also cause a kink
structure in the energy dispersion if one assumes that Eliashberg
function $\alpha^2 F({\bf q},\omega)$ has the same features as
$\chi({\bf q},\omega)$, namely peaked at the wave vector {\bf Q} and
at the Debye frequency $\omega_D$, i.e.  $\omega=\omega_{D} \approx
\omega_{sf}$. By analyzing Fig.~\ref{fermi} it is clear that both spin
fluctuations and electron-phonon coupling can cause a kink structure.
However, in general, one would expect that its position and doping
dependence might be different in both cases. For example, only in the
case of dominant spin fluctuation coupling the kink structure can be
related to INS experiments, i.e.  $\mbox{Im }\chi({\bf Q},\omega)$,
and, furthermore, the kink position is given by $\omega_{kink} \approx
E_{\bf k-Q} + \omega_{sf}(x)$. As discussed earlier, the kink feature
along the antinodal $(0,\pi) \to (\pi,\pi)$ direction results from the
structure in $\phi(\omega)$. Thus, additional structure in
$\phi(\omega)$ due to the electron-phonon interaction (EPI) may also
contribute.  Therefore, the question remains: How to distinguish
between spin fluctuations and phonons as a reason for the kink
formation? To answer this question one has to understand how
consistent are both scenarios with available experimental data.  For
example, as was shown Zeyher and Greco \cite{zeyher}, the value of
electron-phonon coupling extracted from the kink analysis yields the
value of the electron-phonon coupling $\lambda$ and T$_c$ that are too
low to account for high-$T_c$ superconductivity in the cuprates.
Furthermore, assuming that the kink structure arises only from the
EPI, it is difficult to understand the $d_{x^2-y^2}$-symmetry of the
superconducting order parameter and related observed anisotropy of the
kink structure (see Appendix \ref{appb}). Note, only the spin
fluctuation scenario yields $T_c \approx 70$K \cite{mabe,mabed}, a
$d_{x^2-y^2}$-wave order parameter, and a kink feature in qualitative
agreement with experiment.  Also the doping dependence of the kink is
difficult to explain within the phonon scenario.  In contrast to Eq.
(\ref{nodalsf}) one would expect $\omega_{kink} \approx E_{\bf k-Q} +
\omega_{D}(x)$ in the case of electron-phonon coupling.

\begin{figure}[h]
\vspace{0.5ex}
\hspace*{0.3cm}
\centerline{\epsfig{file=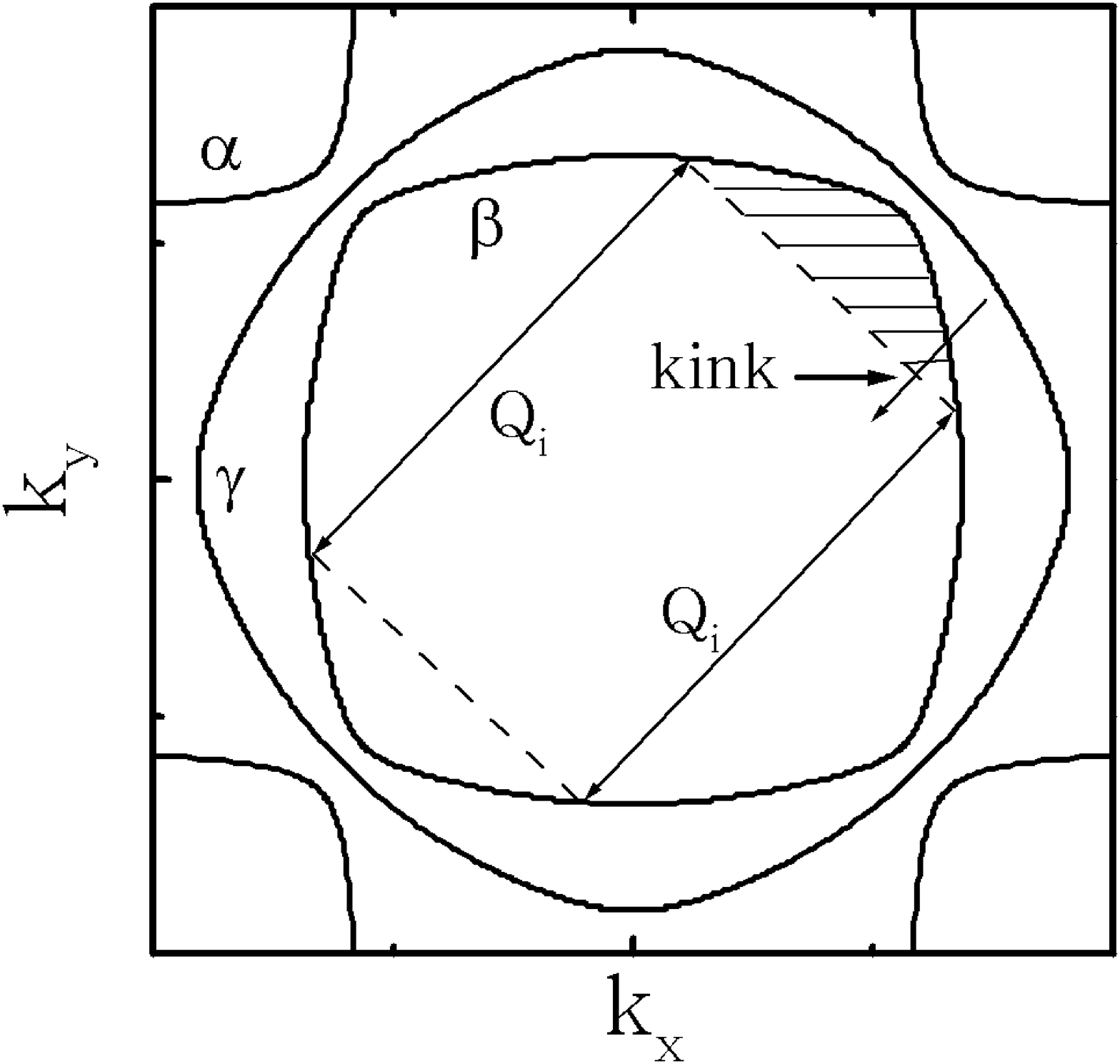,width=6.5cm,angle=0}}
\vspace{0.5ex}
\caption{Illustration of a
possible kink structure in $\mbox{Sr}_2\mbox{RuO}_4$.  The Fermi
surface of $\mbox{Sr}_2\mbox{RuO}_4$ consists of three bands. The
nesting properties of the $\beta$-band yield to a formation of
two-dimensional incommensurate spin fluctuations at ${\bf Q}_i =
(2\pi/3, 2\pi/3)$ and $\omega_{sf} \approx 6$ meV. Therefore the
quasiparticles at the $\beta$-band should be strongly renormalized
due to coupling to spin fluctuations.}
\label{rutenkink}
\end{figure}

{\bf Kink structure in the triplet superconductor
$\mbox{Sr}_2\mbox{RuO}_4$?}
Finally we want to emphasize that the formation of the kink feature
due to spin fluctuations should not be restricted to cuprates. For
example, the quasi-two-dimensional triplet superconductor
$\mbox{Sr}_2\mbox{RuO}_4$ (isostructural to $\mbox{La}_2\mbox{CuO}_4$)
\cite{maeno94} reveals pronounced incommensurate antiferromagnetic
spin fluctuations at the wave vector ${\bf Q}_i = (2\pi/3,2\pi/3)$ and
frequency $\omega_{sf} \approx 6$meV that originates from the nesting
properties of the quasi-one-dimensional $\alpha$ and $\beta$-bands
\cite{sidis,mazin,morr} (see Fig. \ref{rutenkink} for an
illustration).  On general grounds one would expect a kink structure
in the renormalized energy dispersion of the quasiparticles. Although
the correlation effects are weaker in $\mbox{Sr}_2\mbox{RuO}_4$ ($U$
is smaller), and ${\bf Q}_i$ is an incommensurate wave vector, similar
conditions as in cuprates are present.  Note, the kink feature should
occurs at smaller energies than in cuprates due to a lower value of
$\omega_{sf}$ in the ruthenates. Further experimental studies should
test our suggestion.
\acknowledgements
We thank J. Fink, D. Dessau, and D. Fay for helpful discussions.  The
work of I. E. is supported by the 'Alexander von Humboldt' foundation
and CRDF Rec.007.  We further thank the DFG via the SFB 290 and INTAS
(Grant No. 01-0654) for financial support.
\begin{appendix}

\section{Generalized Eliashberg equations}
\label{appa}

\begin{figure}[t]
\centerline{\epsfig{file=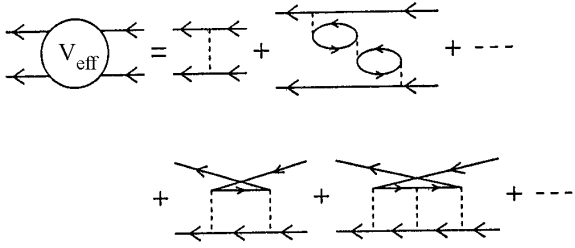,width=8cm,angle=0}}
\vspace{2ex}
\caption{Particle-particle channel of the Bethe-Salpeter equation
for superconductivity due to an effective pairing interaction
$V_{\rm eff}$ entering in Eq. (\ref{flex1}). The solid lines refer
to $G$ and the dashed lines denote the effective Coulomb interaction
$U$ of Eq. (\ref{eq:hubbard}).  Vertex corrections that would yield
to a renormalized coupling strength $U_{\rm eff}$ (as indicated in
Fig. \ref{diagram}) are neglected.  The summation of the
corresponding bubble and ladder diagrams is performed up to
infinity. While in principle it is possible to treat $V_{\rm
  eff}\{\chi\}$ and $G({\bf k},\omega)$ on different levels, we
assume that both quantities are generated by the {\it same}
itinerant quasiparticles.  Note, $V_{\rm eff}$ refers to the
exchange of spin and charge fluctuations yielding a
$d_{x^2-y^2}$-wave instability of the normal state.}
\label{veff}
\end{figure}

The interdependence of elementary excitations with spin excitations
leads to strong self-energy effects.  The corresponding Dyson equation
yields the dressed 2x2 matrix Green's function $G$ in terms of the
bare Green's function $G_0$ and the self-energy $\Sigma$:
\begin{eqnarray}
G^{-1}(k)
& = &
G_0^{-1}(k) - \Sigma(k)
\nonumber\\
& = &
i\omega_nZ(k)\tau_0 - \left[\epsilon(k) +
\xi(k)\right]\tau_3 - \phi(k)\tau_1
\quad ,
\label{A3}
\end{eqnarray}
where $k = ({\bf k},i\omega_n)$.  In the FLEX approximation for the
Hubbard Hamiltonian the self-energy $\Sigma$ is determined by the
following generalized Eliashberg equations:
\begin{eqnarray}
\lefteqn{
\Sigma(k) =
}\nonumber\\
&  &
\sum_{k'}\left[ P_s(k-k')\tau_0G(k')\tau_0 +
P_c(k-k')\tau_3G(k')\tau_3\right] \nonumber\\
& = &
\sum_{k'}\, V_{\rm eff}(k-k')\, G(k')
.
\end{eqnarray}
In order to provide a better understanding of our numerical procedure
we show the corresponding Feynman diagrams for $V_{\rm eff}$ in Fig.
\ref{veff}.  Within RPA the spin and charge fluctuation interaction
are given by
\begin{equation}
P_s = (2\pi)^{-1} U^2 \,
\mbox{Im } (3\chi_s -\chi_{s0})
\quad ,
\label{ps}
\end{equation}
with $\chi_s =\chi_{so}(1-U\chi_{s0})^{-1}$ and
\begin{equation}
P_c = (2\pi)^{-1} U^2
\, \mbox{Im } (3\chi_c -\chi_{c0})
\quad ,
\label{pc}
\end{equation}
with $\chi_c =\chi_{c0}(1+U\chi_{c0})^{-1}$.  Therefore, the kernel
$I$ and the spectral functions of the one-particle Green's function in
Eq. (\ref{flex1}), $A_{\nu}$, read
\begin{eqnarray}
I(\omega,\Omega,\omega') = \frac{f(-\omega')+
b(\Omega)}{\omega+i\delta -
\Omega - \omega'} +\frac{f(\omega')+
b(\Omega)}{\omega+i\delta -
\Omega - \omega'}, 
\label{flex2}
\end{eqnarray}
\begin{eqnarray}
A_{\nu}({\bf k}, \omega) = -\pi^{-1} \mbox{Im }
\left[ a_{\nu} ({\bf k},\omega)
/D({\bf k},\omega)\right],
\label{flex3}
\end{eqnarray}
and 
\begin{eqnarray}
D = [\omega Z]^2 - [\epsilon_{\bf k}^0 +\xi]^2 -\phi^2,
\label{flex3a}
\end{eqnarray}
\begin{eqnarray}
a_0 =\omega Z, \,\,\,\, a_3 = \epsilon_{\bf k}^0 +\xi, \,\,\,\,   
a_1= \phi.
\label{flex3b}
\end{eqnarray}
In Eq. (\ref{flex2}) $f$ and $b$ are the Fermi and Bose
distribution function, respectively. Finally, the bare
susceptibility is calculated from
\begin{eqnarray}
\mbox{Im }\chi_{s0,c0} & = & \frac{\pi}{N} \int_{-\infty}^{\infty} 
d\omega' \left[ f(\omega') -f(\omega'+\omega) \right] \nonumber \\
& & \times \sum_{\bf k} [N({\bf k+q},\omega'+\omega) 
N({\bf k},\omega') \nonumber \\
& & \pm  A_1({\bf k+q},\omega'+\omega) 
A_1({\bf k},\omega') ]
\quad ,
\label{flex4}
\end{eqnarray}
where we assume that the {\it same} itinerant carriers are responsible
for the elemenatry excitations and, at the same time, generate the
spin excitations. In Eq. (\ref{flex4}) we use $N({\bf
  k},\omega)=A_0({\bf k},\omega) +A_3({\bf k},\omega)$, and the real
parts are calculated with the help of the Kramers-Kronig relation. The
subtracted terms in $P_s$ and $P_c$ remove a double counting that
occurs in second order.  Note that $V_{\rm eff}$ in Eq. (\ref{flex1})
is dominated by the exchange of spin fluctuations due to the fact that
the system is in the vicinity of an antiferromagnetic phase
transition, but the above equations still remain valid in the
case where $\chi_c$ becomes more important.

Our numerical calculations are performed on a square lattice with $256
\times 256$ points in the first Brillouin Zone and with $200$ points
on the real $\omega$-axis up to 16$t$ on a logarithmic scale.  Within
our self-consistent procedure the full momentum and frequency
dependence of quantities is kept.

\section{Phonons and ${\bf d}_{\bf x^2-y^2}$-wave superconductivity}
\label{appb}

\begin{figure}[t]
\centerline{\epsfig{file=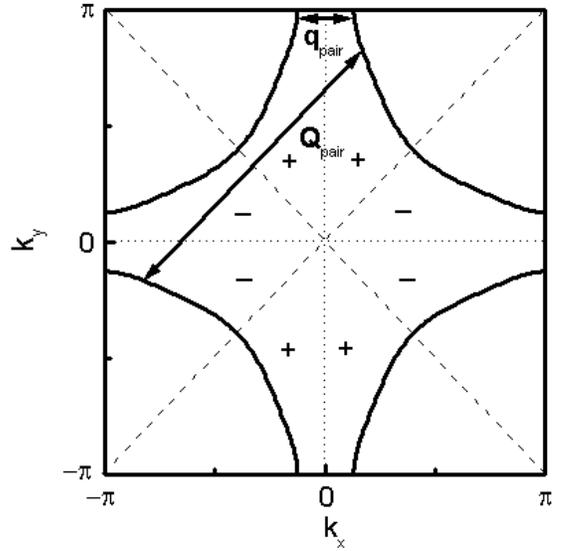,width=8cm,angle=0}}
\vspace{2ex}
\caption{Illustration of $d_{x^2-y^2}$-wave Cooper-pairing for a
fixed frequency $\Omega = \Omega_0 \approx \omega_{sf} \approx
\omega_D$ due to spin fluctuations peaked at momentum ${\bf k} -
{\bf k'} = {\bf q} = {\bf Q}_{\rm pair}$ and due to phonons peaked
at ${\bf q} = {\bf q}_{\rm pair}$. The solid line denotes the Fermi
surface and the dashed line refers to the nodes of the
$d_{x^2-y^2}$-wave order parameter.  The corresponding sign of the
order parameter is also displayed.}
\label{plusphon}
\end{figure}

In this Appendix we analyse how the magnetic mode which is mainly
peaked at $ {\bf q} = {\bf Q}=(\pi,\pi)$ leads to a $d_{x^2-y^2}$-wave
order parameter that is maximal around $(\pi,0)$ and, in particular,
to which extend phonons contribute to this result.  In general, the
generalized Eliashberg equations read after the inclusion of {\it
  attractive} phonons (branch $i$) via their spectral function
$\alpha^2 F_i({\bf q},\Omega)$:
\begin{eqnarray}
\lefteqn{\Sigma_{\nu}^{(i)}({\bf k},\omega) =
} \nonumber\\[1ex]
& & N^{-1}\sum_{\bf k'}
\int_{0}^{\infty} d\Omega \, V_{\rm eff} ({\bf k-k'},\Omega)
- \alpha^2 F_i({\bf k}- {\bf k'},\Omega) \nonumber\\[1ex]
&&\times \int_{-\infty}^{+\infty} d\omega'
I(\omega,\Omega,\omega') \, A_{\nu} ({\bf k'},\omega')
\quad .
\label{flexplus}
\end{eqnarray}
For $\alpha^2 F_i({\bf q},\Omega)$ we employ a Lorentzian in frequency
$\Omega$ around $\Omega_0 \approx \omega_D$ (Debye frequency), and a
normalized form factor $F_i({\bf q})$ peaked at ${\bf q} = {\bf
  q}_{\rm pair}$ as indicated in Fig.  \ref{plusphon}. The spin
fluctuations that are dominating $V_{\rm eff}({\bf q},\omega_{sf})$
are peaked at ${\bf q} = {\bf Q}_{\rm pair}$.

It is instructive to write down the weak-copuling limit of the
$\hat{\tau_1}$-component of Eq. (\ref{flexplus}) that reads ($T=0$)
\begin{equation}
\Delta({\bf k}) = - \sum_{{\bf k'}}\, \frac{\left[ V_{\rm eff}
({\bf q})
- \alpha^2 F_i({\bf q})\right] }{2E_k}\, \Delta({\bf k})
\quad ,
\end{equation}
where again $E_k=\sqrt{\Delta^2({\bf k}) + \epsilon_k^2}$ is the
dispersion of the quasiparticles in the superconducting state.  Note
that the contribution to the pairing potential is repulsive for spin
fluctuations and attractive for phonons, respectively. In the case
where no phonons would contribute to the Cooper-pairing ($\alpha^2
F_i({\bf q}) = 0$), $V_{\rm eff}({\bf q})$ bridges parts of the Fermi
surface where the superconducting order parameter has opposite signs.
This momentum dependence of the pairing interaction is indeed required
for solving Eq. (\ref{flexplus}) and is typical for unconventional
superconductivity. Note that for a repulsive and momentum-independent
pairing potential, $V_{\rm eff}({\bf q}) = const$, no solution
of Eq. (\ref{flexplus}) can be obtained.

How is the kink related to the pairing mechanism? Physically
speaking, the interdependence of elementary excitations that
dominate $V_{\rm eff}({\bf q})$, leads to $d_{x^2-y^2}$-wave
Cooper-pairing as well as to the kink structure as observed by
ARPES experiments. In other words, the quasiparticles around the
hot spots couple strongly to spin fluctuations that leads (a) to
a $d_{x^2-y^2}$-wave order parameter, and (b) the same coupling leads
to the kink in the nodal direction that occurs {\it close to} the
Fermi level where ${\bf Q}_{\rm pair} = (\pi,\pi)$ as indicated in
Fig. \ref{fermi}.

It follows also from Eq. (\ref{plusphon}) that attractive phonons with
a corresponding spectral function $\alpha^2 F({\bf q})$ peaked at
${\bf q} = {\bf q}_{\rm pair}$ contribute {\it constructively} to
$d_{x^2-y^2}$-wave pairing as long as the main pairing interaction is
provided by spin fluctuations.  However, the kink close to the
antinodal points occurs {\it only} below $T_c$ and is a result of
$\phi(\omega)$ that is maximal around $(0,\pi)$.  Therefore, the kink
structure in the antinodal direction is mainly connected to spin
excitations peaked at ${\bf Q}_{\rm pair} = (\pi,\pi)$ and not to the
phonon branch peaked at ${\bf q}_{\rm pair}$.

Note, in the case where no spin fluctuations would be present, i.e.
$V_{\rm eff}({\bf q}) = 0$, the attractive phonon contribution will
cancel the minus sign on the RHS of Eq. (\ref{flexplus}) yielding an
order parameter with $s$-wave symmetry. Thus, we safely conclude that
both, $d_{x^2-y^2}$-wave Cooper-pairing and the anisotropy of the
kink feature in the elementary excitations are hardly to recoincile
within the same physical picture.

\end{appendix}

\begin{references}
\bibitem{remark1} We have shown in Ref. \protect\onlinecite{mabed}
that the calculated Fermi surface corresponds to an optimum
doping concentration $x=0.15$ with the highest $T_c$.
\bibitem{remark2} The value $U=4t$ makes the comparison
with Refs. \protect\onlinecite{mabed} and \protect\onlinecite{manske}
convenient. We obtain similar results for different values of $U$
($3 \leq U \leq 6$).
\bibitem{migdal0} This would not be the case
for electron-phonon mediated Cooper-pairing due to
Migdal's theorem \protect\cite{migdal}.
\bibitem{mabed0} The corresponding phase diagram $T(x)$ of hole-doped
cuprates is shown in Fig. 3 of Ref. \protect\onlinecite{mabed}.
\bibitem{remark3} Note, this estimate of $\omega_{sf}$ corresponds
to the peak frequency in $\mbox{Im }\chi$. While $U$ controls
essentially the height of the peak, its position is only weakly
$\omega$-dependent. The $t$-dependence of $\omega_{sf}$ is reflected
by the fact that nesting gives a pronounced peak in $\mbox{Im }\chi$.
A careful analysis of the Ornstein-Zernicke
expression for the dynamical spin susceptibility (see Eq.
(\ref{mmp})) is given by Moriya and Ueda in Ref.
\protect\onlinecite{moriya}.
\end{references}
\end{document}